# Phonon Transport Controlled by Ferromagnetic Resonance

Chenbo Zhao[1,2], Yi Li[1], Zhizhi Zhang[1], Michael Vogel[1], John E. Pearson[1], Jianbo Wang[2], Wei Zhang[3,1], Valentine Novosad[1], Qingfang Liu[2*], and Axel Hoffmann[1,4*]

[1] Materials Science Division, Argonne National Laboratory, Argonne, Illinois 60439, USA

[2] Key Laboratory of Magnetism and Magnetic Materials of the Ministry of Education, Lanzhou University, Lanzhou 730000, People's Republic of China

[3] Department of Physics, Oakland University, Rochester, MI 48309, USA

[4] Department of Materials Science and Engineering, University of Illinois at Urbana-Champaign, Illinois, IL 61801, USA

**ABSTRACT**

The resonant coupling of phonons and magnons is important for the interconversion of phononic and spin degrees of freedom. We studied the phonon transmission in $LiNbO_3$ manipulated by the dynamic magnetization in a Ni thin film. It was observed that the phonons could be absorbed strongly through resonant magnon-phonon coupling, which was realized by optimizing the interfacial coupling between Ni and $LiNbO_3$. The line shapes of phonon transmission were further investigated considering the magnon-phonon interconversion in the elastically driven ferromagnetic resonance process. The results promote unique routes for phonon manipulation and detection in the presence of magnetization dynamics.

**INTRODUCTION**

Elastically driven ferromagnetic resonance (FMR) is at the core of combining straintronics and spintronics [1-5], which is drawing much attention due to both interesting fundamental physics and potential applications. This includes among others, elastically driven spin pumping [6,7], phonon driven inverse Edelstein effect [8] and field-free magnetization switching [9]. Recently, several studies on magnons-phonons interconversions have emerged [10-14]. In particular, theoretical models were developed in order to provide a microscopic understanding of how magnon-phonon interaction influences the damping and transport of magnons [13,15]. The mechanisms for the coupling between phonons and magnons in ferromagnetic materials include, magnetostriction [1-9,16] and spin-rotation coupling [14,17,18]. The latter mechanism is a manifestation of the Einstein–de Haas and Barnett effects corresponding to the transfer of the angular momentum between spin and mechanical degrees of freedom [19-21]. In other words, the transfer between magnon- and phonon-angular momentum can be manipulated by magnetization dynamics [15,19,21-24]. These earlier research works shed light on the opportunity for phonon manipulation [25] and detection using magnetization dynamics. However, a remaining challenge is to accurately measure the changes of the phonon systems due to the magnon-phonon interaction [4]. Therefore, we focus here on how the propagation of phonons can be modulated via ferromagnetic resonance.

Based on surface acoustic wave (SAW) magneto-transmission measurements, the interaction of SAWs and ferromagnetic thin films has been studied experimentally by several groups [1,2,5,8,26]. However, important aspects of the interaction mechanism still remain to be resolved. An important parameter to characterize the interaction mechanism between the SAWs and ferromagnetic thin films is the linewidth of the transmission power. In general, the linewidth of the transmission power obtained from phonon driven FMR is much larger than that in cavity FMR [1,2,26]. In order to fit the broadening of the linewidth of the transmission power obtained from phonon driven FMR, the posteriori damping constant used is several times larger than the damping values obtained from cavity FMR experiments [1,2]. At the same time, the significant broadening of the transmission power line shape contains important information about the magnon and phonon coupling [4], *e.g.*, angular momentum transformation, during phonon driven FMR. Therefore, the phonon transport properties in the presence of phonon driven FMR also provide an effective approach for exploring the angular momentum transformation [11,27] via the coupled phonons and magnons.

In the present work, we have investigated the dependence of the phonon transport on magnetization dynamics in Ni/LiNbO$_3$ hybrid heterostructures. We observe the simultaneous existence of dips and peaks in the transmission power of the surface acoustic waves (SAW) device, which we interpret by considering both phonon attenuation and generation [28,29]. By fitting the theory [28,29] to the phonon transmission power, the value of the FMR field, resonance linewidth and the exchange stiffness parameter of the magnon system can be estimated. The obtained resonant linewidth of the SAW driven FMR is very similar to those from waveguide FMR experiment. This suggests that the phonon driven FMR linewidth broadening is mainly due to magnon-phonon interaction, instead of the nonuniform excitation fields induced by SAW. Our findings thus generate new opportunities for phonon control and detection of magnetization dynamics.

**EXPERIMENT AND SET UP**

Figure 1(a) shows a schematic illustration of the magnon-phonon interaction in our SAW driven FMR measurement. In these devices, we excite, at microwave frequencies, the SAW using interdigitated transducers (IDT) on a 127.86º Y-X cut $LiNbO_3$ substrate with the thickness of 500 μm. The SAW then drives subsequently FMR in the Ni film. The optical images of the $Ni/LiNbO_3$ hybrid device with a 50-nm thick Ni rectangular film is shown in Fig. 1(b) and (c). The IDTs of 50-nm thick Al were fabricated by using maskless photolithography and electron-beam evaporation. The finger width and the spacing of the IDT were both 3 μm, launching SAWs with a wavelength $\lambda_{SAW}$ of 12 μm. The Ni film was deposited via DC sputtering using 10 W power, and an Ar pressure of $3\times10^{-3}$ Torr. The length of the delay line $L$ was set at $L = 438$ μm, and the Ni film was deposited in the center of delay line. In order to observed phonon driven FMR, a DC magnetic field $H$ is applied in a direction with an angle $\theta$ with respect to the SAW wavevector $k_{SAW}$ set to $\theta =60º$ and $\theta =90º$, respectively. During phonon driven FMR measurements $H$ is varied from -1.1 to 1.1 kOe. As shown in Fig. 1(d), the reflection parameters of the device exhibit one peak at $f_0$= 325 MHz, measured using a vector network analyzer (VNA). The velocity of this SAW mode can be determined as $v_{SAW} = 3912$ m/s, according to the following relation between the velocity $v_{SAW}$ and frequency $f_0$ of SAW: $v_{SAW} = \lambda_{SAW} f_0$. The measured $v_{SAW}$ is typical for the velocity of a Rayleigh SAW [30].

To separate the SAW signals and the electromagnetic wave, the time-gating function of VNA was used [1,26]. Figure 1(e) shows the transmission parameters of SAW signal for the device with gating on (centered at 145 ns and spanning over 45 ns) [1,26]. The SAW odd harmonics from $3^{th}$ to $13^{th}$ have been launched and the highest harmonics is located at 4.23 GHz. Because the $9^{th}$ and $13^{th}$ harmonics are weak, this work focuses on the $3^{th}$, $5^{th}$, $7^{th}$ and $11^{th}$ harmonics to investigate the phonon transport during the coupling of magnon and phonon relevant to the Rayleigh SAW. Some of the device with Ni films were annealed at 400 °C for 30 minutes in vacuum, and the as-grown and annealed devices both were measured to study the phonon transport properties [see Fig. 1(a)].

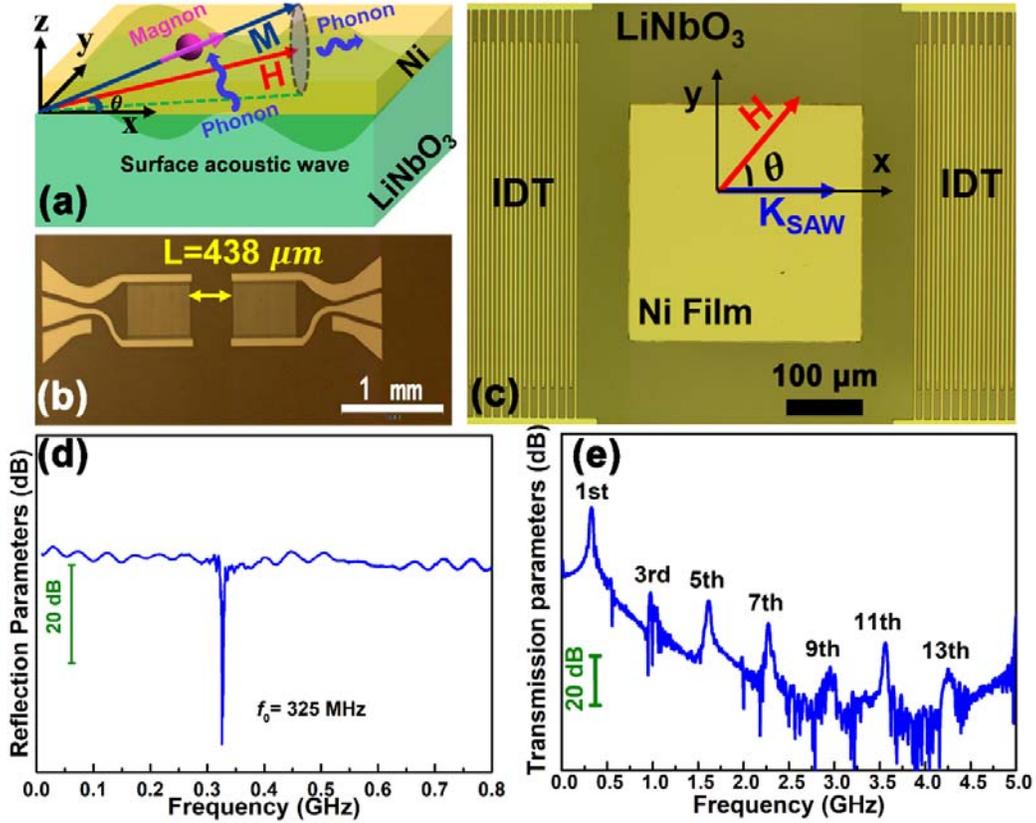

Figure 1 (a) shows schematic illustration of the magnon-phonon interaction during SAW driven FMR; (b) and (c) show optical images of the Ni/LiNbO3 hybrid device with 50-nm thick Ni rectangular film; (d) shows the reflection parameters for the device with $H$ = 1000 Oe applied at $\theta$ = 60º; (e) shows the transmission parameters of SAW signal with gating on (centered at 145 ns and spanning over 45 ns) $H$ applied at $\theta$ = 60º.

**RESULTS AND DISSCUSSION**

The SAW is generated and detected by electromagnetic wave using a pair of spatially separated IDTs on the LiNbO$_3$ crystal. Coupling with the magnon system (Ni films) is achieved during the SAW propagation in between the two IDTs. Phonon-driven FMR can be characterized by the transmission parameter $S_{21}$ using the VNA with time gating function[1,26]. To enhance the signal-to-noise ratio, the transmission parameter $S_{21}$ of SAW was converted into transmission power $P$. Figure 2(a-f) show the colormap of magnetic field $H$ dependence ($H$ is applied at $\theta = 60°$) of the normalized transmission power $P$ of the SAW signal: (a-c) for an as-grown device at $5^{th}$, $7^{th}$ and $11^{th}$ harmonics, respectively; (d-f) for an annealed device at $5^{th}$, $7^{th}$ and $11^{th}$ harmonics, respectively. The color codes represent intensity of transmission power of SAW, with red indicating maximum transmission. The dotted green lines represent the Kittel formula, which is based on waveguide FMR measurements for an extended Ni film grown on LiNbO$_3$. The colormap data shows that the intensity of SAW exhibits a minimum near zero field for the as-grown device, but an obvious attenuation near the FMR field is observed for the annealed devices. This shows that the annealing treatment improves the resonant coupling between the LiNbO$_3$ and the ferromagnetic Ni material. The enhancement of the resonant magnon-phonon coupling can be attributed to improving the interface of the magnon-phonon system [31] via the thermal annealing treatment. This magnon-phonon coupling, as indicated by the dispersion crossing, is enhanced due to the strong absorption of phonons at $f = 2.24$ GHz and 3.56 GHz for the annealed device. The attenuation of the SAW transmission power near zero field [Figs. 2(a-c)] and the FMR field [Figs. 2(d-f)] can be attributed to the magnetization switching [1,32] and the FMR absorption [1-5], respectively.

The plots of the magnetic field dependence of the SAW signal peak position are shown in Fig. 3 and illustrate the phonon transport properties. Figure 3(a) and (c) show the magnetic field ($H$) dependence of the normalized transmission power for the as-grown device at $\theta = 60°$ and $90°$, respectively. When $H$ is applied at $\theta = 60°$, the transmission power shows large dips near the zero field region due to the magnetization switching, which has been reported before [1,32]. Notably, peaks for transmission power near the FMR field are also observed [Fig. 3(a)]. These peaks indicate that the SAW transmission intensity increases near the FMR field. When $H$ is applied at $\theta = 90°$, only the magnetization switching dips remain [Fig. 3(c)]. This shows that for $\theta = 60°$ ferromagnetic resonance is excited. However, there is only the uniform mode that participates in the coupling for $\theta = 90°$.

We find that the annealing treatment for the device can markedly tune the phonon transport. Figure 3(b) and (d) show the $H$ dependence of the transmission power for the annealed device at $\theta = 60°$ and $90°$, respectively. The dips at the FMR field and the peaks above the FMR field in the transmission power spectra are both observed at $\theta = 60°$ [Fig. 3(b)] simultaneously, but they are not present when $\theta = 90°$ [Fig. 3(d)]. This suggests that both attenuation and amplification of the phonon transport is possible due to the magnon-phonon interaction. The $H$ position of the peaks [Fig. 3(a)] and dips [Fig. 3(b)] for as-grown and annealed devices are plotted in

Figs. 3(e) and (f), respectively, as well as the expected behavior based on the Kittel formula. The $H$ positions are in good agreement with the Kittel formula, which indicate that the phonons transport is manipulated by magnetization dynamics during FMR.

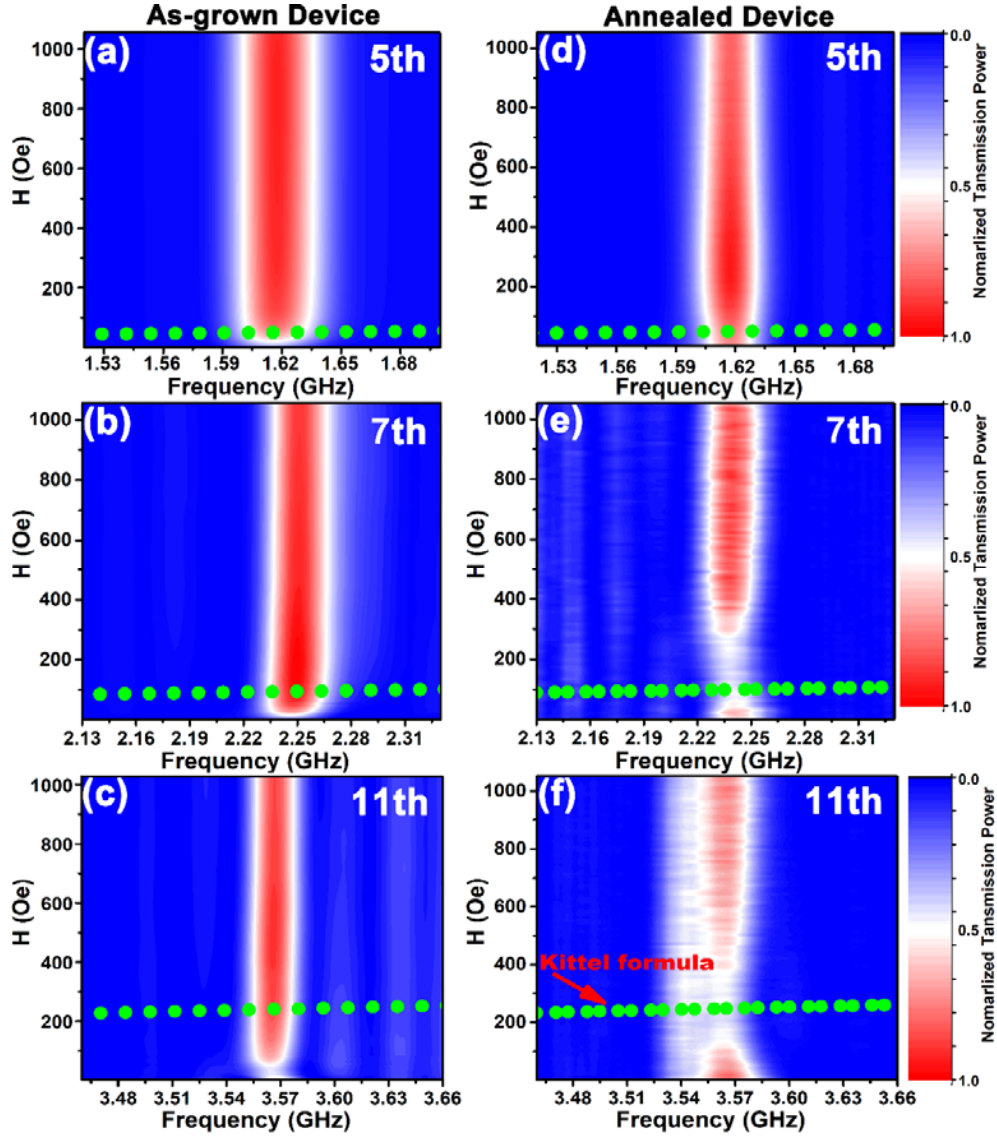

Figure 2: Colormap of magnetic field dependence of the normalized transmission power of SAW signal for the as-grown and the annealed devices at different SAW harmonics with $H$ applied at $\theta = 60°$: (a-c) for the as-grown device at the 5th, 7th and 11th harmonic; (d-f) for the annealed device at the 5th, 7th and 11th harmonic. The dotted lines represent the Kittel formula, which come from waveguide FMR experiment for the full Ni film. The color codes represent intensity of transmission power of SAW, with red indicating maximal transmission.

To model the phonons transport, we take into consideration both phonon attenuation and generation [28,29] manipulated by the magnetic dynamics, during the SAW-driven FMR. Angular momentum interconversion between magnon and phonons is related to the Einstein–de Haas effect [14,18] and the Barnett effect [21]. Due to the magnon-phonon interconversion during FMR, phonons will redistribute the angular momentum and energy between magnons and phonons [15,17], and thus macroscopic mechanical rotation can be strengthened, leading to an enhancement of the transmission of the SAW. The change of the SAW transmitted power ($E$) related to the phonon attenuation and generation during FMR can be rewritten as [28,29]:

$$E = E_0 \frac{v_t}{c} \frac{B_2^2}{C_{44}} \frac{|h_{eff}^\perp|}{\Delta H^2} (1+p^2)^{-\frac{1}{2}} [1+(p+\beta)^2]^{-\frac{1}{2}} \exp\left[\frac{-\frac{\eta}{2}}{1+(p+\beta)^2}\right] \left[1 + \exp\left(\frac{-\eta_0 H}{2}\right)\right] (1),$$

where,

$$p = \frac{H - H_r}{\Delta H} \quad (2),$$

$$\eta = \frac{2\pi f B_2^2 L}{C_{44} v_t M_s \Delta H \mu_0} \quad (3),$$

$$\beta = \frac{2A\rho(2\pi f)^2}{C_{44} M_s \Delta H \mu_0} \quad (4),$$

and the $H$ dependence for the line shape of the transmission power is determined by the parameters $p$, $\eta$ and $\beta$. Equation (1) corresponds to the conventional microwave transmission through a Ni film [28,29], but employs instead of an externally applied real microwave magnetic field the effective *ac* magnetic field $h_{eff}^\perp$ due to the magnon-phonon interaction defined in Eq. (5) below. The parameters in Eqs. (1-4) are: $v_t$ = 3912 m/s is the Rayleigh acoustic wave velocity, $c$ = 3×10$^8$ m/s is the speed of light, $B_2$ = 8.7×10$^6$ N/m$^2$ is the magnetoelastic parameter of Ni, $C_{44}$ = 1.22×10$^{11}$ N/m$^2$ is the elastic modulus of Ni, $\Delta H$ is the linewidth of the FMR, $L$ = 12 μm is characteristic length related to the SAW wavelength, $H$ is the external magnetic field, $H_r$ is the FMR field, $M_s$ = 4.7×10$^5$ A/m is the saturation magnetization, $A$ is the exchange stiffness parameter, $\rho$ = 8900 kg/m$^3$ is the density of Ni, $f$ the frequency of the SAW and $\mu_0$ is the permeability of vacuum. The last term in formula (1) is the phonon scattering due to the viscosity and thermo-conductivity [28,33,34].

The field angle dependence ($H$ at θ = 60º and 90º) for the line shape of the transmission power [Figs. 3(a) and (b); Figs. 3(c) and (d)] exhibits the fingerprint for the Rayleigh SAW driven FMR [1,5], which is relevant to the components $|h_{eff}^\perp|$ of the effective ac magnetic field $h_{eff}$ perpendicular to the magnetization [1,2,5]:

$|h_{eff}^\perp| = |h_x \sin\theta - h_y \sin\theta - ih_z| = 2(b_1 - b_2)\varepsilon_{xx0} \cos\theta \sin\theta + 4b_6\varepsilon_{zx0} \cos\theta$ (5),

where $b_1$ and $b_2$ are the longitudinal-type magnetoelastic coupling constants, $b_6$ is the shear-type magnetoelastic coupling constant, $\varepsilon_{xx0}$ and $\varepsilon_{zx0}$ are pure Rayleigh SAW strain components. As can be seen from Eq. (5), the value of $|h_{eff}^\perp|$ is zero at θ =90º. Thus, the SAW cannot excite the FMR in that configuration. Therefore, both the dips and peaks due to the phonon attenuation and generation observed for θ =60º [Figs. 3(a) and (b)] are related to the FMR and vanish at θ =90º [Figs. 3(c) and (d)]. Unlike the asymmetry of the dips for opposite $H$ orientations in previous results [1,5,8], the dips are nearly symmetry in Fig. 3(b). The asymmetry has been found being proportional to the magnetoelastic coupling constant $(b_1 - b_2)$ [5]. Therefore, when the values of $b_1$ and $b_2$ are close to each other, the asymmetry becomes less apparent. This also indicates that the shear-type magnetoelastic coupling ($b_6$) is dominant over the longitudinal-type magnetoelastic couplings ($b_1$ and $b_2$) for our device.

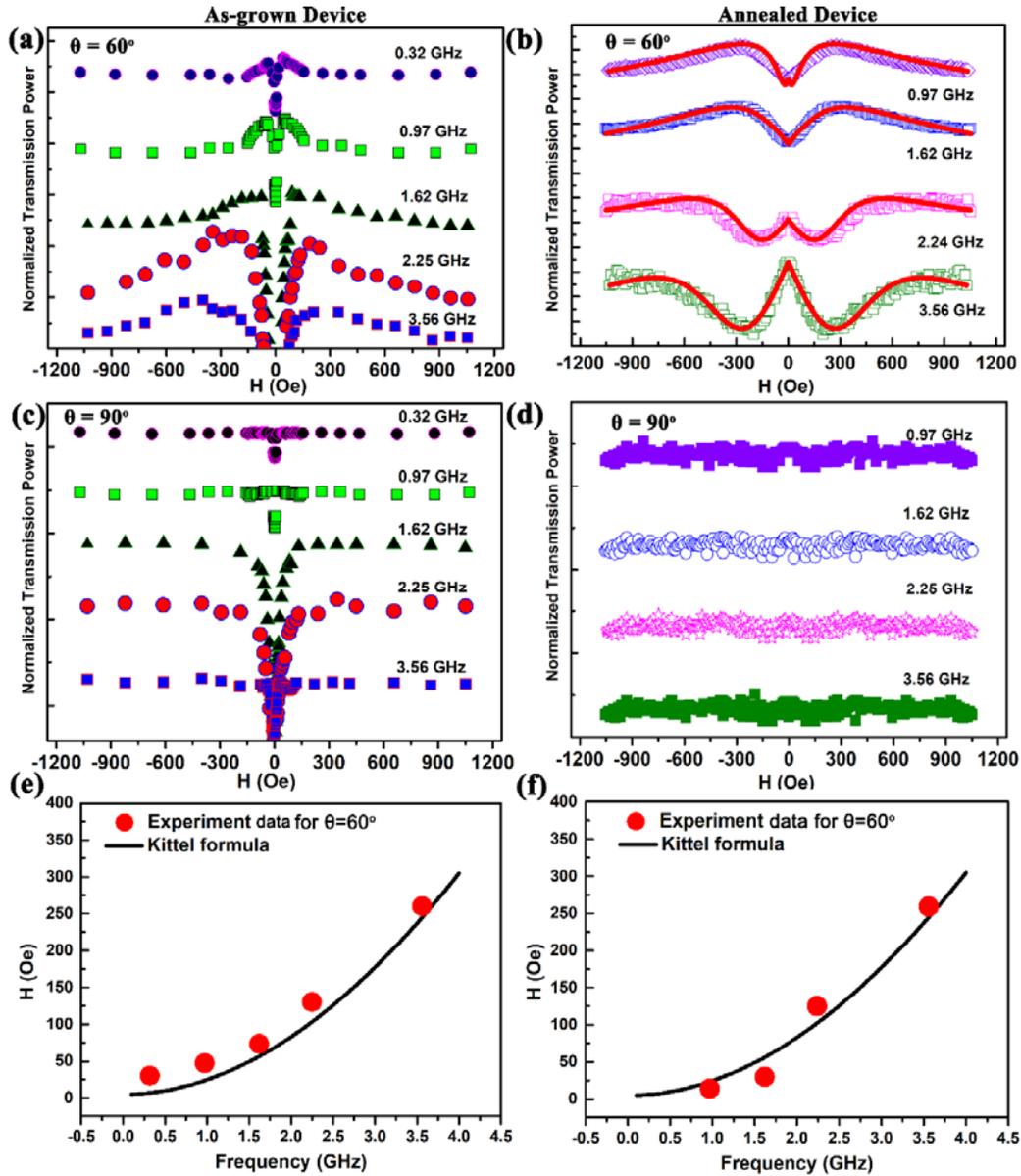

Figure 3: (a-d) Magnetic field $H$ dependence of the transmission power [the symbols are experimental data, and the solid lines are fit to theory using Eq. (1)]: (a) as-grown at $\theta = 60°$; (b) annealed device at $\theta = 60°$; (c) as-grown at $\theta = 90°$; (d) annealed device at $\theta = 90°$. (e) shows the $H$ position of the transmission power peaks at angle $\theta = 60°$ for the as-grown sample; (f) shows the $H$ position of transmission power dips at angle $\theta = 60°$ for the annealed device. The Kittel formula is determined from waveguide FMR experiments for full Ni film.

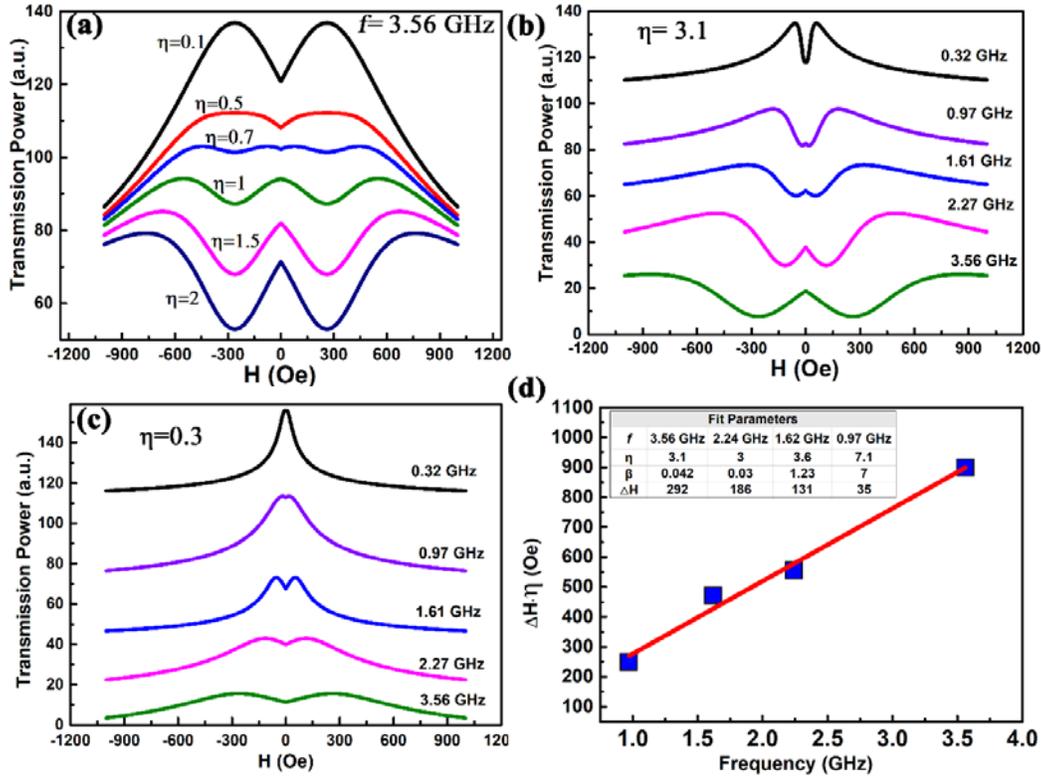

Figure 4: (a-c) Theoretical calculation results of the magnetic field $H$ dependence of the transmission power at different frequencies using Eq. (1): (a) With different $\eta$ at $f$=3.56 GHz; (b) and (c) with values of $\eta$=3.1 and $\eta = 0.3$ at different frequencies. The transmission power are relative values and the cures are shifted with respective to each other. (d) Linear fitting of $\eta$ and $\Delta H$ using $\Delta H\eta = kf$ ($k$ is a constant), where $\eta$ and $\Delta H$ are obtained from theory fits to the line shape of the transmission power at different frequency $f$ for the annealed device using Eq. (1), as shown in Fig. 3(b).

The experimental results for the annealed device are shown in Fig. 3(b). The line shape of the transmission power can be fitted using Eq. (1), and the fit parameters $\eta$, $\beta$ and $\Delta H$ are shown in the table in Fig. 4(d). Since only $\Delta H$ and $\eta$ change with $f$, the expression of $\eta$ can be simplified as $\Delta H\eta = kf$ ($k$ is a constant), resulting in a good fit for $\eta$ and $\Delta H$, as shown in Fig. 4(d). This suggests that the model describes the phonon transport in these SAW devices very well. By putting the fitted value of $\beta$ into Eq. (4), the exchange stiffness parameter of Ni film is estimated as $A = 0.54 \times 10^{-6}$ ergs/cm ($f = 3.56$ GHz) and $0.64 \times 10^{-6}$ ergs/cm ($f = 2.24$ GHz). These obtained values for $A$ are slightly smaller than $A = (0.76 \pm 0.03) \times 10^{-6}$ ergs/cm obtained from elastic small-angle neutron scattering for nanocrystalline Ni film [35]. At a lower frequency, $f = 0.97$ GHz, the anomalous value of $\eta$ and $\beta$ obtained from theory fits can be attributed to the nonuniform magnetization distribution, which also results in large inhomogeneities of the magnetization precession.

However, for the transmission power of the as-grown device [Fig. 3(a)], the experimental data cannot be fitted using Eq. (1) due to the strong attenuation given by the magnetization switching. According to the materials-specific parameters for Ni and the FMR parameters at $f = 3.56$ GHz determined from an independent waveguide-FMR experiments, $\eta$ is calculated to be $\eta = 3.1$ using Eq. (3). This calculated value of $\eta$ is consistent with fits of the data to Eq. (1). We neglect the phonon scattering due to the viscosity and thermo-conductivity and set $\eta_0 = 0$ [28]. The theoretical calculations using Eq. (1), i.e., the $H$ dependence of the transmission power with different values of $\eta$ at $f = 3.56$ GHz, are shown in Fig. 4(a). The transmission power is shown using relative values and the cures are shifted with respective to each other. The calculations show that the dips at $H = H_r$ are increasingly obvious with the increase of $\eta$, which means the phonons are strongly attenuated at the ferromagnetic resonance field. The peaks related to the phonon generation at $H = H_r + \Delta H \sqrt{2\eta - 1}$ ($\eta > 1/2$) become more obvious with the decrease of $\eta$. When $\eta \leq 1/2$, only the peaks at $H = H_r$ are observed. Figures 4(b) and (c) show theoretical calculations of the $H$ dependence of the transmission power with values of $\eta = 3.1$ and $\eta = 0.3$ at different frequencies, respectively. It can be seen that the line shape of the transmission power has a good agreement with the annealed [Fig. 3(b), $\eta > 1/2$] and as-grown device [Fig. 3(a), $\eta \leq 1/2$, except the dips near zero field due to magnetization switching], respectively. The changes in $\eta$ related to magnetoelastic coupling [Fig. 3(a) and (b)], can be attributed to improvement the interface of the magnon-phonon [31] system by the annealing treatment.

Furthermore, the linewidth $\Delta H$ of the transmission power can be obtained to be $\Delta H = 325$ Oe and 500 Oe at $f = 2.24$ GHz and 3.56 GHz from the transmission power dips plotted in Fig. 3(b), which are approximately 1.7 times larger than the values $\Delta H$ = 185.3 Oe and 290.6 Oe obtained from waveguide FMR experiments. These values are consistent to the theory fitting results ($\Delta H$ = 186 Oe and 292 Oe) using Eq. (1). Therefore, compared to the FMR measurements, any broadening effect coming from nonuniform excitation fields induced by SAW [1,2,26] can be ignored. The increase of linewidth can mainly be attributed to the magnon-phonon interaction, e.g., angular momentum transformation [15,23]. These results show that the coupling between elastic and magnetic degrees of freedom open additional channels for information interconversion between phononic and magnonic components.

**CONCLUSION**

In conclusion, phonon transport properties during the phonon driven ferromagnetic resonance has been investigated. The ferromagnetic resonance is driven acoustically, since no external *rf* magnetic field is applied to the ferromagnet. Rather, a purely internal *rf* magnetic field arises due to magnetoelastic coupling between the surface acoustic wave elastic strain field and the ferromagnet. Annealing of the sample results in increase interfacial magnon-phonon coupling, and thus enables tuning of the line shape of transmission power. Considering both the phonon attenuation and generation simultaneously during the phonon driven FMR, the phonon transport properties and line shape of transmission power can be explained in a quantitative fashion. By analyzing the phonon transmission power, the magnetization dynamics of the magnon system can be detected. We also demonstrate that the broadening effect of the transmission power line shape during phonon driven FMR can be mainly attributed to the interaction of magnons and phonons, instead of the nonuniform excitation fields induced by surface acoustic wave.

**ACKNOWLEDGEMENTS**
This work was performed at the Argonne National Laboratory and supported by the Department of Energy, Office of Science, Materials Science and Engineering Division. The use of the Centre for Nanoscale Materials was supported by the US. Department of Energy (DOE), Office of Sciences, Basic Energy Sciences (BES), under Contract No. DE-AC02-06CH11357. Chenbo Zhao acknowledges additional financial support from the China Scholarship Council (no. 201806180105) for a research stay at Argonne. The author thanks José Holanda for uesfull disscussions.